# Predicting the spatiotemporal diversity of seizure propagation and termination in human focal epilepsy

**Timothée Proix**[a, b, c]**, Viktor K. Jirsa**[d]**, Fabrice Bartolomei**[d]**, Maxime Guye**[e]**, and Wilson Truccolo**[a, b, c, 1]

[a]Department of Neuroscience, Brown University, Providence, Rhode Island, United States of America; [b]Institute for Brain Science, Brown University, Providence, Rhode Island, United States of America; [c]Center for Neurorestoration & Neurotechnology, U.S. Department of Veterans Affairs, Providence, Rhode Island, United States of America; [d]Aix Marseille Univ, INSERM, INS, Inst Neurosci Syst, Marseille, France; [e]Aix Marseille Univ, CNRS, CRMBM UMR 7339, Marseille, France

Recent studies have shown that seizures can spread and terminate across brain areas via a rich diversity of spatiotemporal patterns. In particular, while the location of the seizure onset area is usually invariant across seizures in a same patient, the source of traveling (2-3 Hz) spike-and-wave discharges (SWDs) during seizures can either move with the slower propagating ictal wavefront or remain stationary at the seizure onset area. In addition, although most focal seizures terminate quasi-synchronously across brain areas, some evolve into distinct ictal clusters and terminate asynchronously. To provide a unifying perspective on the observed diversity of spatiotemporal dynamics for seizure spread and termination, we introduce here the Epileptor neural field model. Two mechanisms play an essential role. First, while the slow ictal wavefront propagates as a front in excitable neural media, the faster SWDs' propagation results from coupled-oscillator dynamics. Second, multiple time scales interact during seizure spread, allowing for low-voltage fast-activity (>10 Hz) to hamper seizure spread and for SWD propagation to affect the way a seizure terminates. These dynamics, together with variations in short and long-range connectivity strength, play a central role on seizure spread, maintenance and termination. We demonstrate how Epileptor field models incorporating the above mechanisms predict the previously reported diversity in seizure spread patterns. Furthermore, we confirm the predictions for synchronous or asynchronous (clustered) seizure termination in human seizures recorded via stereotactic EEG. Our new insights into seizure spatiotemporal dynamics may also contribute to the development of new closed-loop neuromodulation therapies for focal epilepsy.

epilepsy | neural fields | seizure termination | Epileptor | tractography

**B**rain dynamics during epileptic seizures present various micro and macroscopic spatiotemporal structures at different seizure stages. For patients with drug-resistant epilepsies, better understanding the mechanisms underlying these spatiotemporal patterns during seizure initiation, propagation and termination is crucial to improve treatment methods including resective surgery of seizure onset zones (1–3) and new therapeutic approaches such as seizure prediction or abatement by electrical stimulation (4, 5).

Understanding and modeling the complex mechanisms underlying the many aspects of seizure spatiotemporal dynamics is a difficult task, as different spatial and temporal scales interact. In addition, as new recording technologies have become available and studies have provided more detailed description of the dynamics of seizure propagation, maintenance and termination, a diversity of apparently contradictory phenomena has emerged. Different models have tackled some of these components, such as the spatiotemporal dynamics of seizure propagation (3, 6–8), the source and direction of SWDs during seizures (6, 9, 10), or the mechanisms supporting seizure onset and offset (11, 12). Here, we focus on two main related issues in the context of seizures that evolve into ictal spike-and-wave discharges (SWDs). First, while seizures propagate slowly to connected areas with speeds on the order of 1 mm/s (13, 14), ictal waves in the form of SWDs propagate orders of magnitude faster (100-1000 mm/s) (6, 10, 15). Furthermore, recent studies report apparently contradictory results regarding the source of fast propagating SWDs. Smith et al. (2016) report that slow ictal wavefronts are the moving source of SWDs. In contrast, a study by Martinet et al. (2017) recently supports that the source driving the fast SWDs remains stationary at the initial seizure onset area. Second, seizure termination has long been characterized as a quasi-synchronous event across the recruited brain areas (6, 10, 11). It is clear, however, that not all seizures terminate synchronously across the brain. In many cases, seizures may evolve into different clusters or regions of activity (7), with ictal activity in each cluster terminating quasi-synchronously, but with long termination delays between the clusters spanning tens of seconds (Fig. 1).

In this article, we introduce a unifying neural field model that explains the diversity of previously observed phenomena including seizure initiation, propagation across local and large-scale brain networks, and termination. To do so, we extend the Epileptor neural mass model (12), which is a canonical model for seizure temporal dynamics, into a neural field model. Neural fields are population-level activations, which are coarse-grained in time and space (16), and capture the spatiotemporal evolution of collective state variables such as synaptic or firing rate activity of neuronal populations. The novel Epileptor field equations incorporate local homogeneous and heterogeneous long-range connectivity. While the homogeneous connectivity represents short-range connections which are currently not resolved by tractography methods, heterogeneous connectivity can be derived from diffusion magnetic resonance imaging (MRI) to build whole-brain models. Two main properties of the Epileptor field model allow us to predict and account for the controversial issues regarding seizure spread and termination described above. First, the propagation of the slow ictal wavefront and fast SWDs are supported by two different mechanisms, i.e. wavefronts in excitable neural media and coupled-oscillator dynamics, respectively (17). While changing the seizure onset area requires changing the model parameters, the source of the SWDs can dynamically move or remain stationary in the initial seizure

[1]To whom correspondence should be addressed. E-mail: wilson_truccolo@brown.edu

1–10

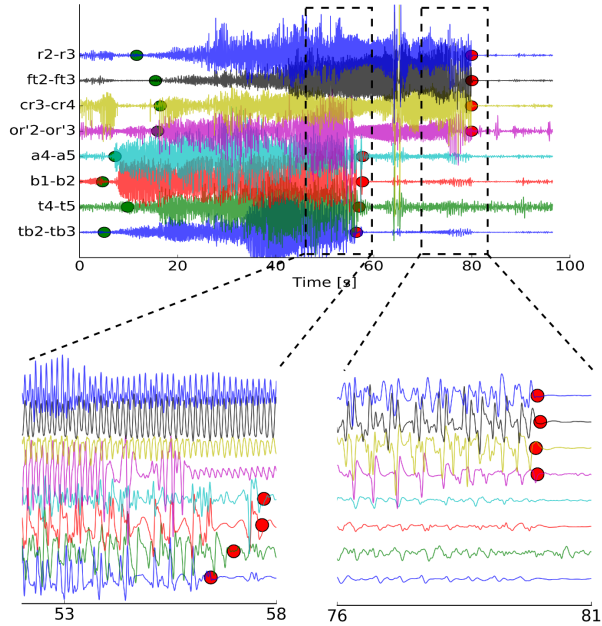

**Fig. 1.** Example of a seizure of patient FB recorded via SEEG electrodes. The top plot shows the full extent of the seizure. Green (red) points indicate seizure channel-wise onset (offset) as determined by visual inspection. Large delays between different brain areas can be observed between different channels both for seizure onset and offset. The bottom left (right) plot magnifies the seizure termination for the first (second) cluster of recorded brain areas. Channels where the seizure ended simultaneously show coherent spike-and-wave activity. Colors denote the corresponding channels in the top and bottom plots.

onset area, depending on the neural field excitability and on coupled-oscillator phase reorganization during the seizure. Second, we show that although the slow propagation of seizures and faster SWDs are supported by two different mechanisms, they are intrinsically related. Specifically, while the dynamics of wave propagation in excitable neural fields determine in part the source and direction of SWDs, the latter contribute to drive the types of spatiotemporal patterns during seizure termination, without requiring the intervention of a global parameter change. In addition, we also show that ($> 10$ Hz) low-voltage fast-activity (LVFA), resulting from the multiple time scale interactions in the neural field model, can hamper the propagation of the slow ictal wavefront. These dynamics, together with variations in short and long-range connectivity strength, play a central role in seizure spread, maintenance and on whether a seizure terminates synchronously or in the form of asynchronous clusters. We demonstrate our predictions in a cohort of 13 epileptic patients, using SEEG recordings and patient-specific tractography obtained from diffusion MRI.

## Results

**Epileptor field model.** We introduce a neural field model with heterogeneous connections based on a neural mass model, called the Epileptor model, originally formulated for the temporal dynamics of the local field potential (LFP) in focal epileptic seizures [12]. To reproduce LFP temporal dynamics, the Epileptor field model (see Materials and Methods) includes neural dynamics with three different time scales related to: (i) the transition between interictal and ictal periods (carried by slow changes in the permittivity variable $s(t)$), (ii) the emergence of LVFA oscillations (carried by the first neural population on a fast time-scale, variables $u_1(t)$ and $u_2(t)$), and (iii) 2-3 Hz SWDs (carried by a second population on an intermediate time-scale, variables $q_1(t)$ and $q_2(t)$). The slow permittivity variable captures slowly evolving physiological process in the brain, such as, for example, changes in the extracellular concentration of different ions [18–22], metabolism [23, 24], and tissue oxygenation [25, 26]. The Epileptor ability to trigger seizures autonomously is controlled by an excitability parameter ($x_0$). For $x_0 > -2.1$, the Epileptor triggers seizures autonomously and is said epileptogenic.

To extend the Epileptor neural mass model to a neural field model, we introduced a spatially homogeneous local coupling between the populations with fast and intermediate time scales, using spatial convolutions with local connectivity kernels (see Materials and Methods). The state variables of the Epileptor field model now depend of both time ($t$) and continuous space ($x$). For simplicity, the coupling function was chosen to be a Heaviside firing rate function. To model seizure dynamics across distal brain areas, we examined a set of two neural fields coupled via long-range heterogeneous connectivity and coupling kernel (see Materials and Methods). Fig. 2 shows a simulation of the two connected Epileptor field models. Although the Epileptor field model can spontaneously transition in and out of seizures (Fig. S1), for convenience, we here trigger seizures through local stimulation at the center of one of the Epileptor fields. Once initiated, the seizure then slowly propagates through an expanding wavefront recruiting adjacent areas from the interictal into the ictal state. In turn, the second Epileptor field is recruited after a longer delay, displaying a similar propagating ictal wavefront. As the seizure evolves, SWDs emerge, propagating with a much higher speed than the slow ictal wavefront. Importantly, the origin and directions of the propagating SWDs can either change through the seizure (Fig. 2), or alternatively, the SWD sources may remain spatially stationary as shown later. Finally, the seizure terminates almost synchronously within each Epileptor field separately. Nevertheless, as in actual seizures (Fig. 1), there can be a substantial termination delay between the seizure offset times between the two fields. In the next sections, we demonstrate how the proposed Epileptor field model accounts for the diversity in spatiotemporal dynamics during seizure spread, maintenance and termination.

**Seizures propagate slowly through brain regions.**

*Ictal wavefront propagation in excitable neural media.* When a seizure initiates in a localized region of the field, this region excites adjacent non-recruited territories: a slow traveling wavefront solution is formed (Fig. 3A) and propagates in the excitable medium. The existence of this slow propagating wavefront solution depends on the local coupling in the fast variable $u_1$ of the Epileptor field model (Eq. 1 in Materials and Methods). The mechanism of this front propagation can be visualized in a phase space plot representing the fast variables $u_1$ and $u_2$ (Fig. 3B). As a seizure starts in a region, the local coupling function will excite the neighboring Epileptors from a stable interictal state into an oscillatory ictal state. The ictal wavefront solution is thus generated by the propagation of the excited activity from the stable fixed point to the oscillatory limit cycle (Fig. 3B, blue lines). Importantly, the slow wavefront solution exists even when the permittivity variable ($s(x, t)$) is constant throughout the Epileptor field model. The permittivity variable values is in part controlled by the excitability parameter ($x_0$). Thus, spatial variation in the excitability parameter is not necessary to obtain a moving ictal wavefront. We also note, as shown above in Figure 2, that a stage-like recruitment from one brain region to another with large delays can be obtained as well by introducing heterogeneous connections. This will be discussed in more detail in the next sections.

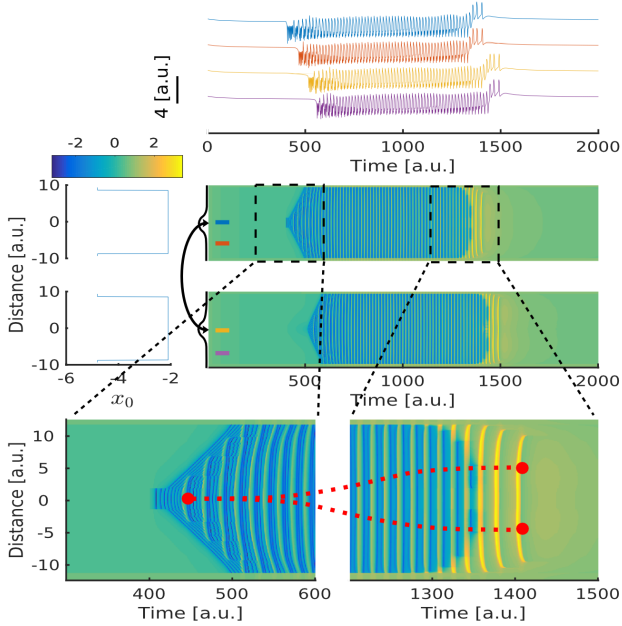

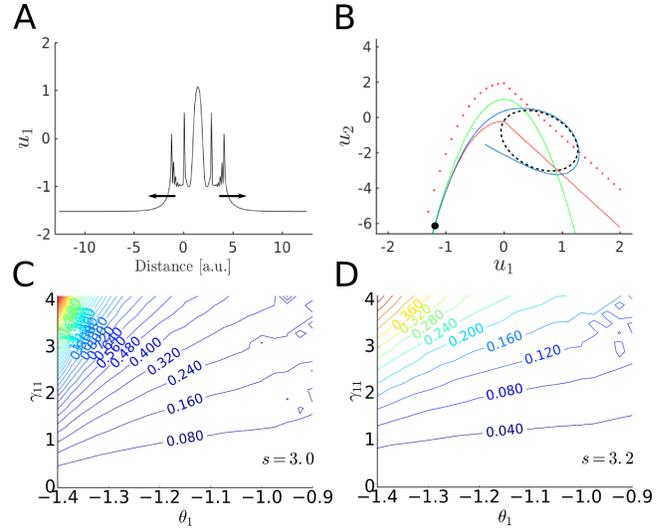

**Fig. 2.** The Epileptor field model reproduces multiscale features of spatiotemporal seizure dynamics: slow ictal wavefront, fast SWD waves with moving sources, and quasi-synchronous termination within distinct neural fields. Two distant Epileptor field models were simulated (middle plot). The first field (upper field) includes the seizure onset area, while the second field (lower field) represents a more distal brain area. Spatial values for excitability parameter ($x_0$) and simulated spatiotemporal activity are shown on the left and right middle plot respectively. For simplicity, we chose a constant spatial profile for $x_0$, except for the regions close to the field boundaries, which represent brain regions with low excitability that are not recruited into the seizure. The center of the two neural fields were coupled via heterogeneous connection as indicated by the curved arrow. The connection kernels (black lines) were centered at the indicated locations and had a scalar multiplicative strength $\gamma_{het,12} = 0.15$. The seizure started in the upper Epileptor field model, slowly propagated throughout the upper field, then eventually recruiting the other Epileptor field model, where it slowly propagated as well. The horizontal colorbar indicates the level of activity for the middle and lower plots. Time series from four spatial locations (indicated by colored bars) are shown in the top plot (line colors are the same as the corresponding colored bars in the middle plot). The lower plot shows a zoomed in view of the spatiotemporal activity at seizure onset (left) and offset (right). The seizure ended synchronously in each field, with a few SWDs propagating after seizure offset. Red dots mark the source location of the propagating SWDs at seizure onset and offset. As the seizure evolved, the source of SWDs changed along with the slow ictal wavefront across each field.

**Fast oscillations at seizure onset hamper seizure propagation.** The slow propagation of epileptic focal seizures has been observed in people with pharmacologically resistant epilepsy (13, 14) and in vitro models (27). It has been attributed to an inhibitory restraint that hinders epileptic activity from propagating faster. As shown above, this slow ictal wavefront propagation is reproduced by the Epileptor field model. To identify the mechanisms constraining the speed of propagation, we reduced the Epileptor field model to a two-dimensional system ($u_1, u_2$) using averaging methods (28). We obtained the speed of propagation of the ictal wavefront by using a semi-analytical method known as the shooting method (SI Text and Fig. S2). The speed of propagation of the front solution was computed as a function of the local coupling strength ($\gamma_{11}$) and the threshold of activation ($\theta_{11}$) for the Heaviside firing rate function in the Epileptor field model (Eq. 1 in Materials and Methods, Fig. 3C and D) for different values of the permittivity variable ($s$). As shown in Figure S3, the semi-analytical shooting method and the numerical simulations of the full system led to consistent results. To correctly evaluate the speed of propagation, the shooting method must include the two variables ($u_1$ and $u_2$) of the fast population, in order to allow for fast oscillations (LVFA) to emerge on the ictal wavefront. The further reduction of the equations by averaging the activity of $u_2$ over time via averaging methods (28) prevented LVFA to occur. Because of the non-linearity of the Heaviside function, such averaging resulted in higher coupling strength and wavefront speeds being one order of magnitude higher (SI Text and Fig. S4). Thus, fast oscillations are a critical element in slowing down the speed of propagation. Indeed, compared to the temporal average of oscillations, oscillation minima led to periodic subthreshold values in the Heaviside firing rate function, therefore periodically setting the effective values of the excitable coupling to zero, and therefore reducing the propagation speed of the ictal wavefront. On the other hand, oscillation maxima do not affect the coupling because of the Heaviside coupling function being flat above threshold. This mechanism stays valid for a non-linear sigmoidal firing-rate function. Although LVFA oscillations have been associated with inhibitory activity at seizure onset [22, 29] and are consistent with the inhibitory veto hypothesis [13], our model points to a more general dynamical mechanism for how LVFA oscillations can effectively hinder seizure propagation. The dynamical mechanism is more general in the sense that other sources of LVFA oscillations, not necessarily requiring the intervention of surround inhibition (inhibitory veto), can lead to the same hampering effect.

**Fig. 3.** Slow traveling ictal wavefront for seizure propagation in the Epileptor field model. (A) Plot of a traveling front solution in the fast population. (B) Phase space for the fast population variables ($u_1$ and $u_2$). In the interictal state, the system is on the left stable fixed point (black dot) at the crossing of the cubic-linear nullcline $\dot{u}_1 = 0$ (full red line) of Eq. 1 and the squared nullcline $\dot{u}_2 = 0$ (green line) of Eq. 2 (Materials and Methods). After seizure onset, excitation through local coupling moves the red nullcline of neighboring Epileptors upward (dashed red line) until disappearance of the left stable fixed point, resulting in the activity (blue line) to jump to the right stable limit cycle (black dashed line). (C) Speed of propagation (a.u.) of the ictal wavefront as a function of the threshold of activation $\theta_1$ and the connectivity strength $\gamma_{11}$ with the permittivity variable $s = 2.95$. The speed of propagation is slow if the threshold of activation $\theta_1$ is higher than the value of the fast population activity ($u_1$) in the interictal state. (D) Same conventions as (C) for $s = 3.2$. The speed of propagation remains slow for other values of the permittivity variable.

## Emergence of coupled-oscillator dynamics across recruited brain areas.

***Coupled-oscillator dynamics accounts for the fast propagation of ictal spike-and-wave discharges.*** The phenomenon of slow ictal wavefront propagation examined above can be understood as wave propagation in excitable media. In contrast, the transition into oscillatory states characterized as ictal SWDs brings additional features belonging to coupled-oscillator systems. Ictal SWDs emerge during the seizure and travel rapidly (100-1000 mm/s [6, 10]) in comparison to the slow ictal wavefront (Fig. 4A). In the Epileptor field model, this is achieved by the local coupling in the second population (variable $q_1$). To understand the mechanisms underlying the fast propagation of SWDs, we used averaging methods [28] to isolate the second population from the other variables. We expressed the second population equations as a function of a variable $K$, whose value is constant under the averaging approximation and depends on the slow permittivity variables and the average activity of the fast population ($u_1$, SI Text). The variable $K$ represents the instantaneous values of slowly evolving variables during the seizure, such as the extracellular concentration of potassium. Simulations of the full model indicated that during the seizure, the variable $K$ takes values that push the second population into an oscillatory regime. The activity of the second population therefore behaves as a chain of coupled oscillators (Fig. 4B). In this case, and in contrast to the slow wavefront propagation in excitable media, the propagation of SWDs originate as phase differences across the different neural oscillators instantiated by the second population in the neural field model. As we show in the next sections, coupled-oscillators propagation support arbitrarily fast SWD propagation and rapid movement of the spatial source of the SWDs, as opposed to wave propagation in excitable media.

We quantified the propagation speed of the SWDs and compared it with the propagation speed of the slow ictal wavefront supported by the fast population. The speed of propagation of the SWDs during the seizure depends on several parameters, such as the instantaneous phase $\phi$ and the intrinsic frequency of each neural oscillator, differences in phase and intrinsic frequency among the oscillators, as well as the variable $K$ and the coupling strength $\gamma_{22}$ between oscillators. To evaluate the speed of the SWDs, we approximated the dynamics of the neural field by the dynamics of a chain of uncoupled relaxation oscillators. This approximation was reasonable as confirmed by the numerical simulation of the full Epileptor field model (Fig. 4C and D). Next, we obtained the speed of the SWDs as a function of the variable $K$ (Fig. 4C) and initial phase $\phi_0$ (Fig. 4D), while fixing all of the other parameters (SI Text).

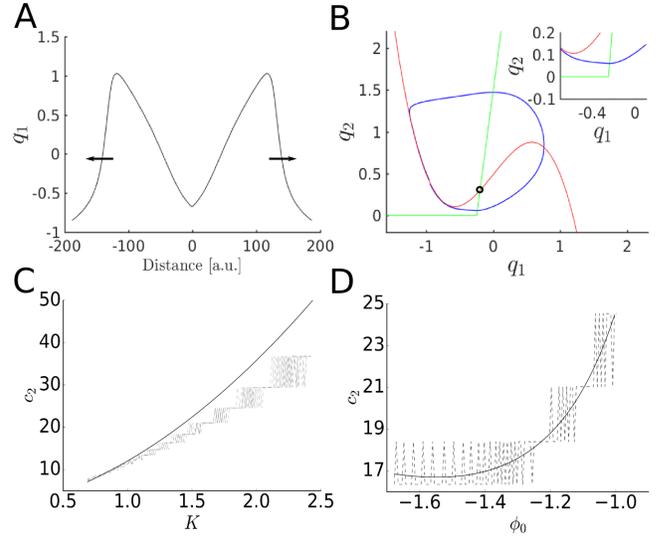

**Fig. 4.** Fast traveling spike-and-wave discharges in the Epileptor field model. (A) Pulse solution for the second population (see also SI Text). The spatial extension of the Epileptor field model is shown larger than in previous figures for a better visualization of the pulse solution. (B) Phase space for the second population variables ($q_1$ and $q_2$). The variable $K$, representing slowly evolving processes (e.g. extracellular potassium concentration) during the seizure, is high enough during the seizure so that the stable fixed point of the system disappears through a SNIC bifurcation leaving only one unstable fixed point, resulting in oscillatory activity. Red: nullcline for the variable $q_1$ ($\dot{q}_1 = 0$). Green: nullcline for the variable $q_2$ ($\dot{q}_2 = 0$). Blue: trajectory of the pulse solution. Black circle: unstable fixed point. (C) Speed of SWDs ($c_2$) in population 2 as a function of the variable $K$, i.e. for fixed values of the permittivity variable and activity of the fast population. The speed of SWDs is two order of magnitude higher than the slow ictal wavefront propagation and can vary substantially according to the values of variable $K$ during the seizure. The black line indicates the prediction from the analytical solution based on a chain of uncoupled relaxation oscillators, while the dashed line indicates the speed obtained by direct numerical integration of the full neural field model, for different values of the coupling strength ($\gamma_{22}$). (D) Speed of SWDs ($c_2$) in population 2 as a function of its initial phase ($\phi_0$) for $K = 1.5$. The speed of SWDs varies according to the instantaneous phase of the different oscillators. Same conventions as in (C).

The speed values we obtained are two orders of magnitude higher than for the speed of propagation of the ictal wavefront, in agreement with previously reported experimental data. Interestingly, the coupling strength does not influence much the propagation speed (Fig. 4C and D).

***Source and direction of spike-and-wave discharges propagation.*** The source and propagation directions of ictal SWDs remains a controversial issue, with apparently contradicting data supporting either a moving source consisting of the slow ictal wavefront itself [10] or a spatially stationary source at the seizure onset area [6]. The Epileptor field model shows that both cases are possible. Specifically, the location of the source is defined as the site in the neural field that first triggers an ictal SWD during the seizure. As stated previously, the location of this initial site depends on several parameters and variables, which may evolve in different ways during different seizures. In particular, for the full Epileptor field model, the variable $K$ changes during the seizure, as shown in Figure 5A. The maximum of this variable $K$ at a given time across the field determines the source of the next SWD. Furthermore, the excitability ($x_0$) affects the time course of $K$ (Fig. 5B). Thus, the location of the maximum of $K$, and therefore the source of the SWDs, changes during the seizure as a function of the excitability of different brain areas. In one dimension,

Proix *et al.*

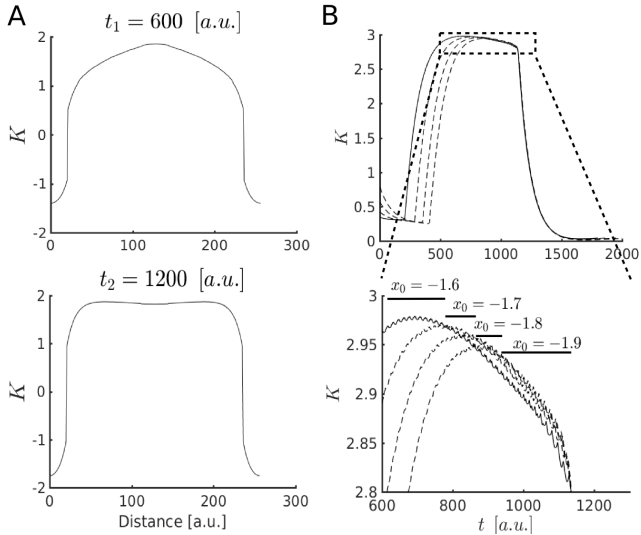

**Fig. 5.** The spatial source of ictal spike-and-wave discharges. The variable $K$, which includes the slow permittivity variable and the average activity of the fast population, determine in part the spatial source of SWDs. The variable $K$ represents the instantaneous values of slowly evolving processes during the seizure and it can evolve differently for different parts of the field during a seizure, thus resulting in a moving source for the SWDs. (A) Evolution of the variable $K$ during a seizure indicated by two different snapshots ($t_1$ and $t_2$). (B) Top plot: the variable $K$ as a function of different excitability levels including $x_0 = -1.6$ (full line) and for values ranging from $-1.9$ to $-1.7$ (dashed lines), at a specific sites of the Epileptor field model. Bottom plot: zoomed in view of the top plot. The maxima of the variable $K$ (black bars), and therefore the sources of propagation, are obtained for different values of $x_0$ at different time points.

the seizure propagates from the source of the SWD in both directions, until it stops propagating or meet another SWD propagating from a different source. In sum, both moving and stationary sources are possible solutions depending on the evolving dynamics. For example, a moving source can occur if the brain region at the origin of the seizure and the recruited areas are of equal epileptogenicity (Fig. 2). On the other hand, given that the variation in the various factors described above can lead to moving sources, it is also not difficult to obtain spatially stationary sources. For example, a spatially stationary source can occur if the brain region at the origin of the seizure is more epileptogenic than the recruited areas (Fig. S1).

**Formation of ictal clusters and stagewise seizure termination.**

*SWDs contribute to quasi-synchronous seizure termination.* Travelling waves and pulses are easily obtained in neural models and have been extensively described (16). Similarly, for an isolated Epileptor field model with only homogeneous connections, one could expect that seizures would propagate as traveling pulses across brain regions, i.e. a slow ictal wavefront would be followed by a slowly propagating offset across the recruited areas. However, electrocorticography (ECoG) and SEEG recordings show that seizures can terminate quasi-synchronously (Fig. 1). We show here that quasi-synchronous seizure termination can occur under certain conditions across an Epileptor field model with homogeneous connections.

To model such synchronous seizure termination within a given ictal cluster, a possible solution is to vary a global pa-rameter for the ictal cluster, which renders the front solution unstable at seizure offset. Instead, we pursued a different mechanism that avoids the assumption of a global parameter. In the original isolated Epileptor neural mass model (12), the seizure ends when the fast oscillatory activity crosses the separatrix between ictal and interictal state through a homoclinic bifurcation. Before seizure termination, these fast oscillations are triggered by traveling SWDs. In turn SWDs are triggered during the seizure by the slow drive of the first population on the second population, i.e. through the temporal convolution $g(u_1)$ acting on $q_1$ (Eq. 4 in Materials and Methods, Fig. 6A). On the other hand, in the Epileptor neural field model, this mechanism was extended spatially through a coupling kernel to trigger the homoclinic bifurcation (Eq. 6 in Materials and Methods). As a consequence, as the field approaches seizure termination, the Epileptors are more likely to cross the separatrix at about the same time with the propagation of a single SWD, resulting in a quasi-synchronous seizure termination (Fig. 2 right lower plot). Therefore, for an isolated Epileptor field model, SWD propagation just before quasi-synchronous termination (small termination delays) also leads to high pairwise correlation values between LFP at different spatial location in the Epileptor field model (Fig. 6B). An important condition for synchronized seizure termination in an Epileptor field model to work is that the recruitment time (related to the propagation of the slow ictal wavefront) is shorter than the seizure length, so that the Epileptors are all close to seizure termination at about the same time. Indeed, it has been observed in ECoG data that total seizure duration is in average $2.3 \pm 0.4$ longer than the propagation duration (14). On the contrary, if the recruitment times are too long, a slowly propagating pulse appears instead, that is, the seizure offset slowly propagates across the field (Fig. S5). In this case, one observes seizure termination as a slowly propagating event as mentioned earlier.

We also note that a few SWDs may still propagate throughout the Epileptor field model after the seizure termination. These SWDs propagate through the recruited brain (Fig. 2 right lower row), and are very similar to the isolated propagating SWDs often observed in ECoG or SEEG recordings. The occurrence and the number of these SWDs after the seizure termination depends on the strength of the coupling function $\gamma_{12}$ in the temporal kernel $g$ (Fig. S6). Nevertheless, we emphasize that the seizure termination would still appear quasi-synchronous if one defines the time of seizure termination as the time of the last SWDs (Fig. 2).

*SWDs fail to propagate between brain regions with large pairwise termination delays.* As seen in Fig. 1, seizures do not always terminate simultaneously across the brain. Instead, many seizures evolve into the formation of ictal clusters. While seizure termination is still quasi-synchronous within each ictal clusters, long termination delays can occur across clusters. According to our model, long termination delays between different brain regions indicate a failure of the SWDs to trigger seizure termination quasi-synchronously in these regions. To reproduce the large delays in seizure recruitment and termination observed between ictal clusters (Fig. 1), here we introduced heterogeneous connectivity between two Epileptor field models (black arrow in Fig. 2 middle row). Such neural field architecture has been previously investigated in detail and named a two-point connection (30, 31).

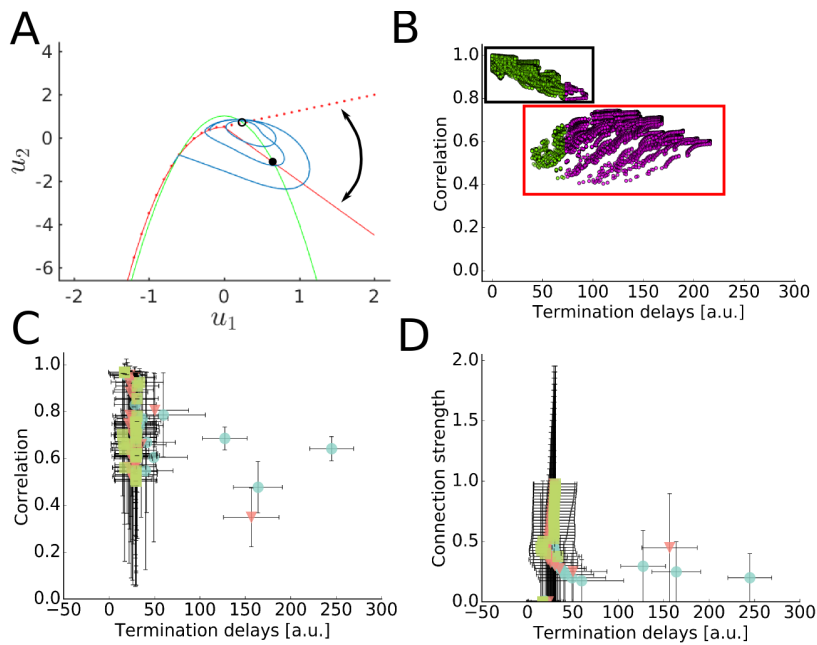

**Fig. 6.** Quasi-synchronized seizure termination results from the fast propagation of ictal SWDs. (A) Phase space for the fast population ($u_1$ and $u_2$) of the Epileptor field model at seizure offset. The slope of the straight part of the red nullcline changes when a SWD occurs, transforming the stable focus (black dot) in an unstable focus (black circle) and allowing for fast oscillations to emerge (blue trajectory) and cross the separatrix in an homoclinic bifurcation. The homoclinic bifurcation results in the termination of the seizure. The curved arrow represents the change in the nullcline before and after a SWD. (B) Pairwise correlations between different sites in the neural field model as a function of the termination delays. Same seizure simulation as in Fig. 2, except that $\gamma_{het,12} = 0.2$. Black (red) square: pairwise correlation for spatial locations in the same (different) Epileptor field models. Lower correlation values are obtained when the pairwise sites are in different fields (red square). Green and violet dot colors indicate different clusters obtained via automated clustering of the termination delays. (C) Mean pairwise correlation between different sites in the Epileptor field model as a function of termination delays for different values of the excitability ($x_{0,2}$) of the second field and of the heterogeneous coupling strength ($\gamma_{het,2}$). Each point corresponds to the mean for each one of the clusters obtained as in (B). Each color/symbol indicate a different excitability value. The vertical and horizontal bars indicate the standard deviations in the corresponding coordinates. (D) Same conventions as in C but showing mean pairwise connection strength, instead.

The Epileptor field model with two-point connection accounts for a seizure starting in a brain region, slowly propagating within this region, and recruiting another brain region with longer delays (Fig. 2). Quasi-synchronous seizure termination can occur between separated Epileptor field models connected via heterogeneous connections, provided that the heterogeneous connection strength $\gamma_{het,12}$ between the two fields is high enough. Since SWDs propagate independently in each field, pairwise correlation values just before seizure termination between LFP sites in two different fields are lower than for one isolated field (Fig. 6B). On the other hand, large termination delays can arise between separated Epileptor field models for lower heterogeneous connection strength $\gamma_{het,12}$ (Fig. 6B).

Next, we examined in more detail the role of the SWDs in triggering quasi-synchronous seizure termination. To quantify if the ability of SWDs to propagate between sites in different fields at seizure termination, we computed the pairwise correlation values just before seizure termination between each spatial coordinate in the two connected Epileptor field models for different heterogeneous connection strength ($\gamma_{het,12}$) and excitability parameter ($x_{0,2}$) values of the second Epileptor field model (see SI Text for details). We used an automated clustering method to identify clusters based on the pairwise termination delays only, i.e. without including the pairwise correlations (Fig. 6B). For each identified cluster, we computed the mean pairwise correlation and mean termination delay (Fig. 6C). The corresponding connections strength for the same clusters are shown in Fig. 6D. As expected, for large termination delays, both the correlation values and the connection strengths are small. Small termination delays can lead to high correlations values when the two site pairs are in the same Epileptor neural field. These high correlations confirm the synchronization of the SWDs in site pairs in the same Epileptor field model. Small termination delays can also be associated with lower correlation values when the delays of recruitment are small and the site pairs are in different Epileptor field models. Additionally, for small termination delays, some clusters can contain pairwise locations both from the same or from separate Epileptor field models (Fig. 6B). In these cases the average connection strength can be low as homogeneous connection strength is ignored. We note that long delays of recruitment are obtained for only a small range of heterogeneous connection strength, and that most seizures have a small delay of recruitment.

**Large seizure termination delays correlate with weak heterogeneous and homogeneous connections in SEEG and tractography data.** We next checked our predictions for the role of SWDs and connection strength in seizure termination in SEEG data recordings. We analyzed 54 seizures from 13 patients (SI Text). Six of these seizures contained large seizure termination time delays between different brain regions. Fig. 7A shows the pairwise site correlation as a function of the termination delay for one seizure in patient FB (same seizure shown in Fig. 1). Pairwise correlations were computed for all electrodes pairs in all patients and seizures. As above, we numerically identified clusters based on the pairwise termination delays only, and we computed mean pairwise correlation values just before seizure

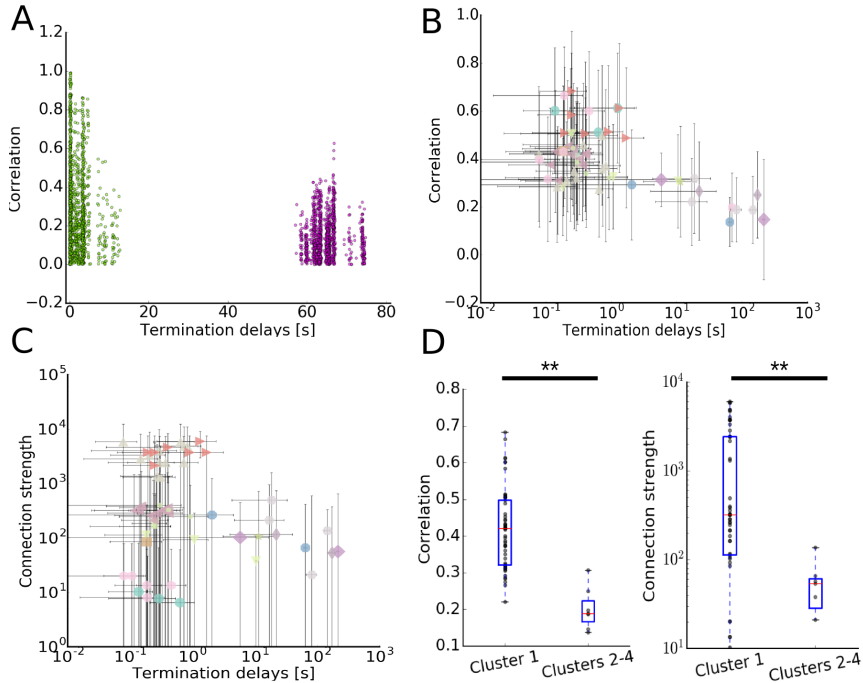

**Fig. 7.** Large differences in seizure termination delays can arise between different ictal clusters involving distant and not well connected regions. (A) Pairwise correlations as a function of the pairwise termination delays for SEEG contacts from an example seizure (Fig. 1). Green and violet dot colors identify clusters based on the pairwise termination delays. (B) Mean pairwise correlation between contacts as a function of the corresponding pairwise termination delays for all contacts, patients and seizures. Each point corresponds to the mean for each one of the clusters obtained as in (A). Each color/symbol indicates a different patient. The vertical and horizontal bars indicate the standard deviation in the corresponding coordinate. Note the log scale in the x axis. (C) Mean pairwise tractography connection strength between contacts as a function of corresponding parwise termination delays. Same conventions as in B. Note the log scale in the x and y axes. (D) Whisker plots for pairwise correlation, and tractography strength for cluster one (smallest termination delays) versus the other clusters (longer termination delays) across patients. Statistically significant differences are indicated (**, P<0.01; Mann-Whitney U-test).

termination and mean pairwise termination delays for each cluster. We found that large termination delays tended indeed to be associated with small correlation values (Fig. 7B), while small termination delays could lead to high and low correlation values.

As suggested by our simulations, we hypothesized that brain regions showing large termination delays are weakly connected via heterogeneous connectivity. To confirm our hypothesis, we processed structural and diffusion MRI data for these patients and used tractography methods to estimate the connection strength between pairs of electrodes as a function of corresponding termination delays (SI Text). Using the same identified clusters as in Figure 7B, we found that large termination delays tended to be associated with small connection strengths (Fig. 7C). We grouped the results for all patients into two groups consisting of small and large termination delays. We found significant differences for correlations and connection strengths between the two groups (Mann-Whitney U-test for intra versus inter clusters, P<0.01 for both measures, Fig. 7D).

To conclude, SWDs ability to propagate at seizure termination is correlated with quasi-synchronous seizure termination and high homogeneous or heterogeneous connection strength, indicating that SWDs plays a role in quasi-synchronous seizure termination. Conversely, low connection strength between two distant brain regions favor the occurrence of long termination delays.

## Discussion

In this article we have introduced a new neural field model that unifies and explains the previously observed diversity in spatiotemporal dynamics of seizure initiation, propagation and termination on different time scales. We have demonstrated how the interplay between temporal and spatial scales leads to: (i) slow propagation of an ictal wavefront whose speed is hampered by fast oscillations (LVFA), (ii) fast propagation of SWDs through coupled-oscillator dynamics, with moving or spatially stationary sources, (iii) the formation of ictal clusters with stagewise recruitment and quasi-synchronous (within clusters) or asynchronous (across clusters) seizure termination. Furthermore, regarding the latter, we confirmed the model's predictions in SEEG and tractography data recorded from epileptic patients that SWD propagation and connection strength correlate with the type of spatial patterns of seizure recruitment and termination. Our results shed new light on the dynamical mechanisms underlying the apparently contradictory diversity of spatiotemporal patterns across different seizures and patients. Altogether, we propose a comprehensive model of spatiotemporal dynamics in epileptic focal seizures, contributing to more accurate individualized patient modeling and better therapeutic interventions.

Mixed connectivities in network architectures are common in large-scale brain models (32, 33). They were first introduced by Jirsa and Kelso (30), and have been used since then in the modeling of resting state networks (31) and evoked potentials induced by electrical stimulation (33). In the neuroinformatics platform The Virtual Brain (34), such neural field architectures are implemented as surface-based modeling approaches, in which a high-resolution cortical surface is equipped with a neural field and homogeneous short-range and heterogeneous long-range connectivity. The latter connectivity is typically obtained from diffusion MRI-derived connectomes and may hold different biophysical mechanisms. In our previous work (3, 35), we modeled heterogeneous coupling between different brain areas via a slow permittivity coupling in different Epileptors. Here, long-range interactions through heterogeneous coupling were modeled via fast populations ($u_1$) at different sites of the Epileptor field model. The two approaches are not incompatible, but rather complementary, as they account for different biophysical mechanisms: fast coupling accounts for synaptic interactions, while slow permittivity coupling accounts for slowly evolving physiological processes such as, for example, extracellular ionic diffusion (e.g. potassium, (20, 36)), spatial buffering of potassium by glial cells (22, 37), as well as increase, via long-range projections in remote regions, of extracellular potassium via increased firing rate (22, 29). In terms of dynamics, while fast coupling explains seizure dynamics such as slow seizure propagation, fast SWDs propagation, and quasi-synchronous seizure termination, slow permittivity coupling can reproduce very large delays of recruitment between distant brain regions, which are difficult to obtain in the model presented here (Fig. 6C and D). However, both models underline the importance of connection strength in seizure spatiotemporal dynamics. In particular, we explained here how brain regions showing low pairwise connection strength are, if recruited, less likely to terminate synchronously and display low pairwise coherence, as experimentally observed by other groups (e.g. (38)).

As stated earlier, recent studies have provided contrasting evidence regarding the emitting source and direction of SWDs. In one study (10), it has been argued that ictal SWDs are emitted by a moving source, which consists of the slowly propagating ictal wavefront. In contrast, (6) have provided evidence for SWDs being emitted by a spatially stationary source consisting of the identified seizure onset area in each seizure. Our results and predictions show that both scenarios (moving and stationary sources) are possible. A critical feature allowing this possibility is the emergence of coupled-oscillator dynamics. Once different sites in the Epileptor field are recruited into the ictal state and transition into oscillatory activity, a network of coupled oscillators is formed. Changes in the location of the source emitting SWDs can result by reorganization of the phases of the neural oscillators. This is a crucial difference between our model and the two previously mentioned studies. In both (10) and (6), the SWDs are only emitted by a small brain region, which can be moving or not. If, during the seizure, the ictal activity in this brain region terminates, for example by electrical stimulation, the seizure will immediately end across the brain. In our model, however, the termination of a seizure might require that ictal activity is abated in all or a much larger set of recruited brain areas. This prediction can be verified using electrical stimulation for seizure abatement in localized areas, and is critical for the development of closed-loop systems for seizure control (39). In particular, our model demonstrate the importance of early seizure detection (e.g. (40)), as this is one of the most favorable times to abate seizures as the recruited area is still small and easier to stimulate. Finally, the emergence of this coupled-oscillator dynamics is to be contrasted with pure traveling wave dynamics in excitable media (17) underlying the slow propagating ictal wavefront. While the propagation dynamics of SWDs can be more complex, seizures tend to propagate as wavefronts in excitable media, in agreement with experimental observations that seizures tend to have a common spatial origin in the same patient.

As a phenomenological model, the Epileptor field model identifies the invariant features that constrain the observed dynamics and may inform the development of detailed biophysical models. It has been suggested that the fast population comprises excitatory neurons, while the population on the intermediate time scales mostly comprises inhibitory neurons (12, 41). Here, we suggest another possibility: small networks of coupled excitatory and inhibitory neurons could be represented by the fast population dynamics, without particularly assigning one variable to a single excitatory or inhibitory population. Instead, the feedback-loop circuitry in which excitatory and inhibitory neurons are embedded allows for the oscillatory behavior captured by the fast population in the Epileptor model. The transition to the SWD activity could be triggered by various factors including, for example, increase of extracellular potassium (22, 29), which can impair inhibitory neuron activity once depolarization block sets in. In the Epileptor field model, the SWD activity increases because of the fast population activity ($u_1$) through a slow temporal convolution ($g$), which increases and then reach a plateau during the seizure. These dynamics affecting SWDs could be instantiated by, for example, changes in extracellular potassium concentration. The SWDs produced by the second population would emerge from the contribution of excitatory neurons only. We also note that here we have focused on epileptic seizures that show SWDs. We hope to address in the future the case of seizures that do not include ictal SWDs, but remain instead within the dynamics of low voltage fast activity oscillations (15, 22, 42, 43)

Patient-specific structural heterogeneous connectivity has recently been used in brain network modeling aiming at personalized medicine, modeling in particular seizure recruitment (1, 3, 44). Other methods have focused on the neural field formulation (6, 9) to capture in particular the SWDs propagation. However, integrating both types of connectivity in a single model can lead to qualitatively different spatiotemporal dynamics than suggested by the combination of dynamics predicted by models using either homogeneous or heterogeneous connectivity (45). By including both homogeneous and heterogeneous connectivity in the same model, we were able to reproduce the diversity of seizure initiation, propagation and termination patterns, and we have demonstrated how different time scales indeed interact together in seizures dynamics. The pitfall of integrating both types of connectivity is the computational cost of large-scale simulations. However, with the increasing computational resources and the advance of neuroinformatic methods such as The Virtual Brain (34), such modeling tools will be accessible to assist clinicians' diagnosis in their daily practice. By improving the accuracy of neural dynamics models across different temporal and spatial scales, we hope to improve the virtualization potential of epileptic brains, the success rate of therapeutical interventions in patients with pharmacologically resistant focal epilepsy, and contribute to the development of closed-loop neuromodulation systems for seizure control.

## Materials and Methods

Detailed discussion of the patient selection, data acquisition and processing, and analytical and computational methods may be found in SI Text.

**Epileptor neural field model.** We extended spatially a five-dimensional model able to reproduce the LFP dynamics of epileptic seizures, know as the Epileptor model (12, 46). The model comprises three different time scales interacting together. The slowest time scale is responsible for leading the autonomous switch between interictal and ictal states, and is driven by a slow permittivity variable. The fastest and intermediate time scales are two coupled oscillators accounting respectively for the LVFA oscillations and SWDs. We introduce here the integral neural field form of the Epileptor model, to account for propagation through short-range connectivity. The five dimensions of the original model are now described as a five-dimensional neural field representing the fast activity ($u_1(x,t), u_2(x,t)$), the spike-and-wave activity ($q_1(x,t), q_2(x,t)$), and the slow permittivity variable $s(x,t)$ at position $x$ and time $t$ as follows:

$$\partial_t u_1 = u_2 - f_1(u_1, q_1) - s + I_1 + \gamma_{11} w_1 * S(u_1, \theta_{11})$$
$$+ \sum_j \gamma_{het,ij} u_{1,j} \qquad [1]$$

$$\partial_t u_2 = y_0 - 5u_1^2 - u_2 \qquad [2]$$

$$\partial_t s = \frac{1}{\tau_0}(4(u_1 - x_{0,j}) - s) \qquad [3]$$

$$\partial_t q_1 = -q_2 + q_1 - q_1^3 + I_2 + 0.002 g(u_1) - 0.3(s - 3.5)$$
$$+ \gamma_{22} w_2 * S(q_1, \theta_{22}) \qquad [4]$$

$$\partial_t q_2 = \frac{1}{\tau_2}(-q_2 + f_2(q_1)) \qquad [5]$$

with

$$g(u_1) = \int_{t_0}^t e^{-\gamma(t-\tau)}(u_1(\tau) + \gamma_{12} w_{12} * S(u_1, \theta_{12})) d\tau \qquad [6]$$

$$f_1(u_1, q_1) = \begin{cases} u_1^3 - 3u_1^2 & \text{if } u_1 < 0 \\ (q_1 - 0.6(s-4)^2) u_1 & \text{if } u_1 \geq 0 \end{cases}$$

$$f_2(q_1) = \begin{cases} 0 & \text{if } q_1 < -0.25 \\ 6(q_1 + 0.25) & \text{if } q_1 \geq -0.25 \end{cases}$$

and $I_1 = 3.1, y_0 = 1, \tau_0 = 2857, \tau_2 = 10, I_2 = 0.45, \gamma = 0.03$. $\gamma$ represents the speed of the temporal integration in Eq. 6. Note that in the original Epileptor $\gamma = 0.01$, the main difference is that SWD propagation is shifted toward the beginning of the seizure as $g$ increases more rapidly (Fig. S7). The parameter $x_0(x)$ represents the excitability of the Epileptor at position $x$ and vary in this article. If $x_0 > -2.91$, the Epileptor is epileptogenic and able to trigger seizures autonomously. Otherwise if $x_0 < -2.91$ the Epileptor stays in a healthy equilibrium state.

Two local coupling functions $\gamma_{11} w_1 * S(u_1, \theta_{11})$ and $\gamma_{22} w_2 * S(q_1, \theta_{22})$ are added in variables $u_1$ and $q_1$ respectively. An additional coupling function $\gamma_{12} w_{12} * S(u_1, \theta_{12})$ spatially extend the action of $u_1$ in the temporal convolution $g$. The term $S(u, \theta)$ is interpreted as a firing rate function, and is chosen here to be the Heaviside coupling function $S(u, \theta) = H(u - \theta)$. Finally, heterogeneous connectivity between the local and distant Epileptor field models $i$ and $j$ respectively is captured by the term $\sum_j \gamma_{het,ij} u_1^j$ is added, with $u_1^j$ corresponding the jth Epileptor field. The spatial convolution $*$ is defined by:

$$w * S(u, \theta) = \int_{-\infty}^{+\infty} w(y) S(u(x-y, t), \theta) dy$$

with $w(y)$ representing the connectivity strength between sites separated by a distance $y$ in a given Epileptor field. $w(y)$ is chosen to be isotropic and homogeneous. Here, we used a Laplacian local connectivity kernel $w(x) = e^{-|x|}/2$. Unless specified otherwise, $\gamma_{11} = \gamma_{22} = \gamma_{12} = 1$.

**ACKNOWLEDGMENTS.** We acknowledge support from the National Institute of Neurological Disorders and Stroke (NINDS), grant R01NS079533 (WT); the U.S. Department of Veterans Affairs, Merit Review Award I01RX000668 (WT); the Pablo J. Salame '88 Goldman Sachs endowed Assistant Professorship of Computational Neuroscience at Brown University (WT). This project received funding from the European Union's Horizon 2020 research and innovation programme under grant agreement No. 720270.

# Supporting Information

**Proix et al.**

## Patient selection and data acquisition

13 drug-resistant patients (6 males, mean age x, range x) with different types of partial epilepsy were selected (Table S1). SEEG electrodes were implanted in the regions suspected to be in the epileptogenic zone. Each electrode has 10 to 15 contacts (length: 2 mm, diameter: 0.8 mm, contacts separation: 1.5 mm). SEEG signals were recorded with a 128 channel Deltamed$^{\text{TM}}$ system (sampling rate: 256 Hz, hardware band-pass filter: 0.16 - 97 Hz). To determine electrode positions, a MRI was performed after electrodes implantation (T1 weighted anatomical images, MPRAGE sequence, TR=1900 ms, TE = 2.19 ms, 1.0 x 1.0 x 1.0 mm, 208 slices) using a Siemens Magnetom Verio 3T MR-scanner. To reconstruct patient specific connectomes (DTI-MR sequence, angular gradient set of 64 directions, TR=10.7 s, TE=95 ms, 2.0 x 2.0 x 2.0 mm, 70 slices, b weighting of 1000 s/mm$^2$, diffusion MRI images were also obtained on the same scanner.

## Data processing

To quantify the strength of connection between electrodes, MRI and dMRI data were processed using SCRIPTS, a processing pipeline to derive individualized cortical surface and large-scale connectivity (40). Cortical and subcortical surface were reconstructed along with a volumetric parcellations using the Desikan-Killiany atlas, with the cortical regions subdivided in four (47) (280 cortical regions and 17 subcortical regions). Head-motions and eddy-currents were corrected in diffusion data. Fiber orientation were estimated with constrained spherical deconvolution, and $2.5 \cdot 10^6$ streamlines were obtained by probabilistic tractography. We used the ACT and SIFT framework to improve reproducibility and biological accuracy (48).

We obtained electrodes positions by coregistering the parcellation with the MRI scan, and assigning each contact to the region containing the most of the reconstructed contact volume.

## Data analysis

Spike-and-wave events were segregated from raw SEEG signals by band-pass filtering between 1 and 10 Hz (fifth-order forward-backward Butterworth filter). For each pair of contacts, the two seconds window before the earliest offset of the two contacts were selected for both contacts (we obtained similar results by using two seconds windows at 2, 4, 6, 8 and 10 seconds before the earliest seizure offset) and correlation computed. Connection strength between pairwise contacts was obtained by the connection strength between associated regions from the parcellation (see Data processing). Both contacts were allowed to be in the same region, as self-connection of the connectivity matrix were not null. We used mean-shift clustering to group pairwise delays and obtain mean and standard deviation of each group.

## Speed of propagation of the seizure - fast population

To evaluate the speed of propagation of the seizure as function of parameters $\theta$, $x_0$, and the variable $s$, we first reduced the Epileptor field model by applying averaging methods. Following (30), we apply averaging methods to neglect the effect of the second population on the first population:

$$\partial_t u_1 = u_2 - u_1^3 + 3u_1^2 - s + I_1 + \gamma_{11} w_1 * S(u_1, \theta) \quad [1]$$
$$\partial_t u_2 = y_0 - 5u_1^2 - u_2$$
$$\partial_t s = \frac{1}{\tau_0}(4(u_1 - x_0) - s)$$

For our choice of connectivity kernel, taking the Fourier transform of Eq. 1 and inverting gives (29):

$$\gamma_{11} S(u_1, \theta) = (1 - \partial_{xx})(\partial_t u_1 + u_1^3 - 3u_1^2 - u_2 - I_1 + s)$$
$$\partial_t u_2 = y_0 - 5u_1^2 - u_2$$
$$\partial_t s = \frac{1}{\tau_0}(4(u_1 - x_0) - s)$$

We introduce the change of variable $\xi = x + c_1 t$:

$$\gamma_{11} S(u_1, \theta) = u_1 u_1''(-3u_1 + 6) + u_1'(-6u_1'(u_1 - 1) + c_1)$$
$$- c_1 u_1''' + u_1^3 - 3u_1^2 - I_1 - u_2 + u_2'' + s - s''$$
$$c_1 u_2' = y_0 - 5u_1^2 - u_2$$
$$c_1 s' = \frac{1}{\tau_0}(4(u_1 - x_0) - s)$$

To obtain the speed of the front, we analyzed the jump-up solution from the interictal state to the ictal state. Using averaging methods, we focused on the fast time scale by setting $s' = 0$. Thus $s = \bar{s}$ is a constant parameter. The system can then be recast as the following system of ODEs:

$$u_1' = v_1$$
$$v_1' = w_1$$
$$w_1' = -\frac{1}{c_1}\big(u_1 w_1(3u_1 - 6) + v_1(6v_1(u_1 - 1) - c_1 + \frac{10u_1}{c_1})$$
$$+ u_2 + \frac{v_2}{c_1} - u_1^3 + 3u_1^2 + I_1 - \bar{s} + \gamma_{11} S(u_1, \theta)\big)$$

$$[2]$$

$$u_2' = v_2$$
$$v_2' = \frac{1}{c_1}(-10v_1 u_1 - v_2)$$

Using this system of equations, we constructed the front numerically by a shooting method. We looked for a heteroclinic solution that links the interictal state fixed point (black full point) to the stable limit cycle (full black cycle). Suppose the speed of the front is given by $c_1^*$, we numerically integrated the system 2 for a given speed $c_1$. Depending if $c_1 < c_1^*$ or $c_1 > c_1^*$, we obtained the blue or the green trajectory in Fig. S2. We now have an optimization problem where we are looking for the speed $c_1^*$ that minimizes the distance between the stable limit cycle and the simulated trajectory, that is the speed $c_1^*$ for which we obtain a front solution. This optimization problem was solved numerically.

**Speed of propagation of the seizure - reduced fast population.**
We here demonstrate that the fast oscillatory activity at the beginning of the seizure reduces propagation speed by deriving the speed of propagation of the seizure when this activity is neglected. We first reduced the Epileptor field model (Eq. [1]) to two degrees of freedom by setting $\dot{u}_2 = 0$, thereby replacing the ictal oscillatory activity by its average (30):

$$\partial_t u_1 = -u_1^3 - 2u_1^2 + 1 + I_1 - s + \gamma_{11} w_1 * S(u_1, \theta)$$
$$\partial_t s = \frac{1}{\tau_0}(4(u_1 - x_0) - s)$$

Proceeding as above, we obtained the following system of ODEs:

$$u' = v$$
$$v' = w$$
$$w' = -\frac{1}{c_1}\big(uw(3u + 4) + v(v(6u + 4) - c_1) - u^3 - 2u^2$$
$$+ 4.1 - \bar{s} + \gamma_{11} S(u, \theta)\big)$$

**Speed of propagation of the SWDs - second population**

For the spatial part of the Epileptor field model which is in the ictal state, the values of $g(u_1)$ and $s$ are such that the recruited Epileptors act as a chain of coupled oscillators (see Fig. 2 in the main text). To isolate population 2 from the full Epileptor field model, we recast the term $0.002g(u_1)$ in Eq. 4 of the main text as $0.002g(u_1) = 2q_3$ with $\dot{q}_3 = -0.01(q_3 - 0.1u_1)$ (12). Using averaging methods along the left branch of the nullcline $\dot{q}_1 = 0$, we can separate the time scales by setting $\dot{s} = 0$ and $\dot{q}_3 = 0$, giving $s = \bar{s}$ and $q_3 = \bar{q}_3$:

$$\dot{q}_1 = -q_2 + q_1 - q_1^3 + I_2 + K(\bar{q}_3, \bar{s}) + \gamma_{22} w_2 * S(q_1, \theta_2) \quad [3]$$
$$\dot{q}_2 = \frac{1}{\tau_2}(-q_2 + f_2(q_1)) \quad [4]$$

with $K$ a constant.

Starting from a stable fixed point $q_i$ for oscillator $i$, the speed of propagation is given by:

$$c_2 = \frac{\Delta x}{\Delta T(q_i, q_j)} = \frac{x_j - x_i}{T_j(q_j) - T_i(q_i)} \quad [5]$$

where $T_i$ is the time for oscillator $i$ to go from fixed point $q_i$ to the knee of the left branch of the cubic nullcline $q_c$ (red nullcline of Fig. 4B of the main text). The coordinates $x_j$ and $x_i$ where chosen as in the discretized problem used for the numerical simulations of the full model. We neglect the effect of coupling on the speed of propagation of the spike-and-waves events ($\gamma_{22} = 0$). This is confirmed by simulation (see Fig. 4C and D of the main text). On the left branch of the nullcline (i.e. $f_2(q_1) = 0$), using averaging methods we can set $\dot{q}_1 = 0$, and substituting in Eq. [4] gives:

$$\Delta T(q_i) = \int_{q_i}^{q_c} \frac{\tau_2(1 - 3q_1^2)}{-q_1 + q_1^3 - I_2 - K(\bar{q}_3, \bar{s})} dq_1 \quad [6]$$

The speed of propagation is then evaluated using Eq. [5] by integrating numerically Eq. [6] for different values of K with fixed initial phase $q_i$ (Fig. 4C), and for different values of the initial phase $\phi = q_i$ with fixed $K$ (Fig. 4D).

This approximation worsen when $K$ increases, as the unique fixed point of the system for $q_1 < -0.25$ is shifted from the knee of the nullcline (Fig. 4B, inset).

**Numerical implementation**

Direct numerical simulations of the Epileptor field model used a Runge-Kutta 4 integration scheme with fast Fourier transform for the calculation of convolutions (49). For the Fig. 2 of the main text, the Epileptor field model activity was set to the fixed point, and a stimulation pulse of strength 1 and spatial width 1.57 was added to Iext at time $t = 400$ for a duration $\Delta t = 10$. Shooting methods used a Nelder-Mead optimization method, and the front solutions were integrated using an Euler integration scheme.

**Table S1. Patient information**

| Patients | Gender | Epilepsy duration (years) | Age at seizure onset (years) | Epilepsy type | Surgical procedure | Surgical outcome | MRI | Histophathology | Side | Number of seizures |
|---|---|---|---|---|---|---|---|---|---|---|
| AC | F | 14 | 8 | Temporal | Sr | III | Anterior temporal necrosis | Gliosis | R | 4 |
| CJ | F | 14 | 9 | Occipital | Sr | III | N | FCD type 1 | L | 3 |
| CV | F | 18 | 5 | Supplementary motor area | Sr | I | N | FCD type 2 | L | 13 |
| ET | F | 23 | 7 | Parietal | Sr | I | FCD SPC | FCD type 2 | L | 10 |
| FB | F | 16 | 7 | Premotor | Sr | II | N | NA | R | 7 |
| FO | M | 45 | 11 | Temporo-frontal | Sr | I | FCD Fr | FCD type 2 | R | 1 |
| GC | M | 5 | 18 | Temporal | Sr | III | Temporopolar hypersignal | FCD type 1 | R | 1 |
| JS | M | 11 | 18 | Frontal | Sr | I | Frontal necrosis (post-trauma) | Gliosis | R | 1 |
| ML | F | 10 | 17 | Temporal | Sr | II | Hippocampal sclerosis) | NA | R | 5 |
| PC | M | 15 | 14 | Temporal | Sr | NO | N | NA | R | 2 |
| PG | M | 29 | 7 | Temporal | Sr | I | Cavernoma | Cavernoma | R | 1 |
| RB | M | 28 | 35 | Temporal | Sr | III | N | Gliosis | L | 1 |
| SF | F | 24 | 4 | Occipital | Sr | NO | PVH | NA | R | 5 |

Th, thermocoagulation; Gk, Gamma knife; Sr, surgical resection; NO, not operated; N, normal; FCD, focal cortical dysplasia; SPC, superior parietal cortex; Fr, Frontal; PVH, periventricular nodular heterotopia; NA, not available; L, left; R, right.

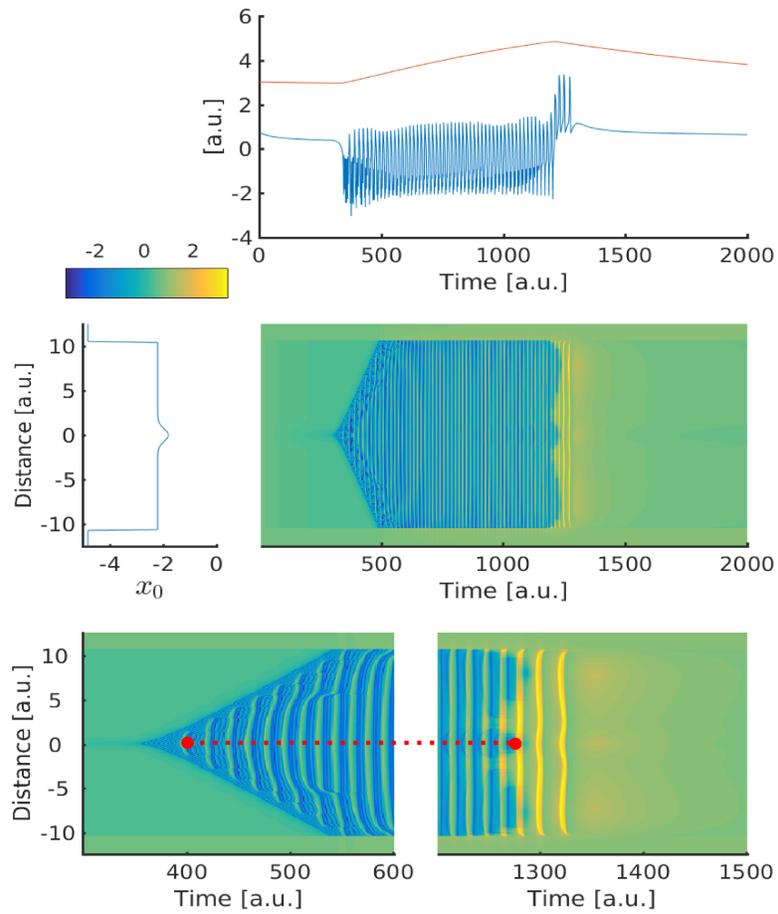

**Fig. S1.** The Epileptor field model can spontaneously transition in and out of seizures, reproducing features of seizure spatiotemporal dynamics. The transitions occur without stimulation, but by changes in the spatial values of the excitability parameter $x_0$. Note that the location of the source of the SWDs stays stationary during the seizure. A single Epileptor fields is simulated (middle row). Spatial values for excitability parameter and simulated spatiotemporal activity are shown on the left and right middle row respectively. The seizure starts in the central location of the Epileptor field model where the excitability is higher, and slowly propagates throughout the field. Time series at the zero location (where the seizure first starts) are shown in the top row for the field activity ($-u_1 + q_1$, in blue) and the slow permittivity variable ($s$, in red). The lower row shows a zoomed-in view of the spatiotemporal activity at seizure onset (left) and offset (right). The seizure ends synchronously in the field, with a few spike-and-wave events propagating after seizure offset. Red dots and the dashed line mark the source location of the propagating SWDs at seizure onset and offset.

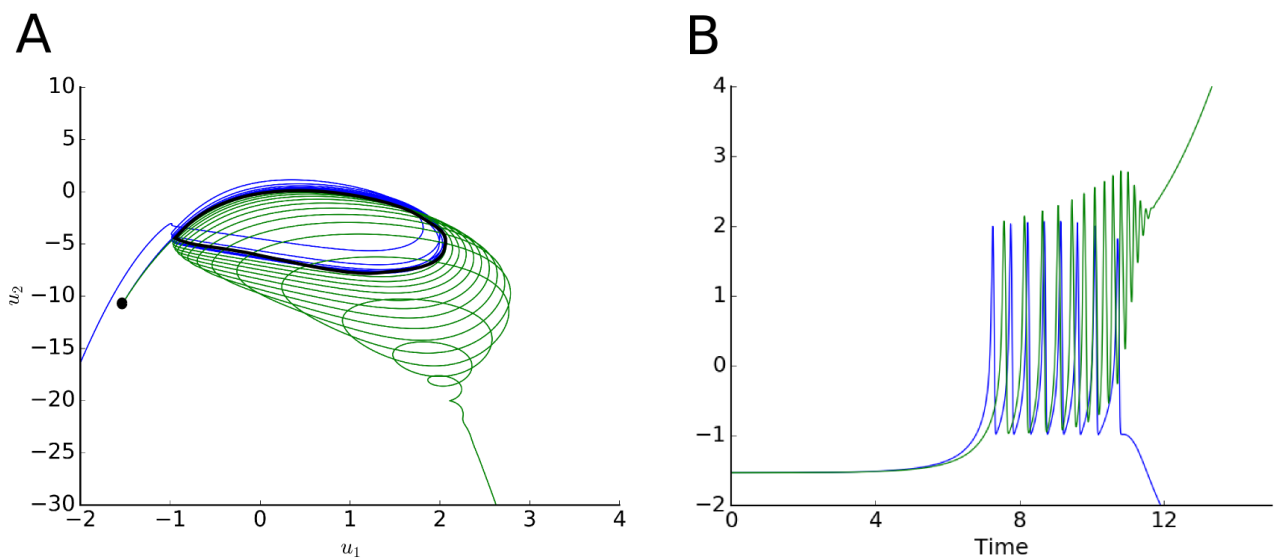

**Fig. S2.** Shooting method to compute propagation speed of the fast population. (A) Phase space for the fast population. Black dot: interictal stable fixed point. Black closed curve: stable limit cycle. Simulated trajectories are shown for $c_1 < c_1^*$ (blue) and $c_1 > c_1^*$ (green), with $c_1^*$ the speed of the front. (B) Corresponding time series for $c_1 < c_1^*$ (blue) and $c_1 > c_1^*$ (green).

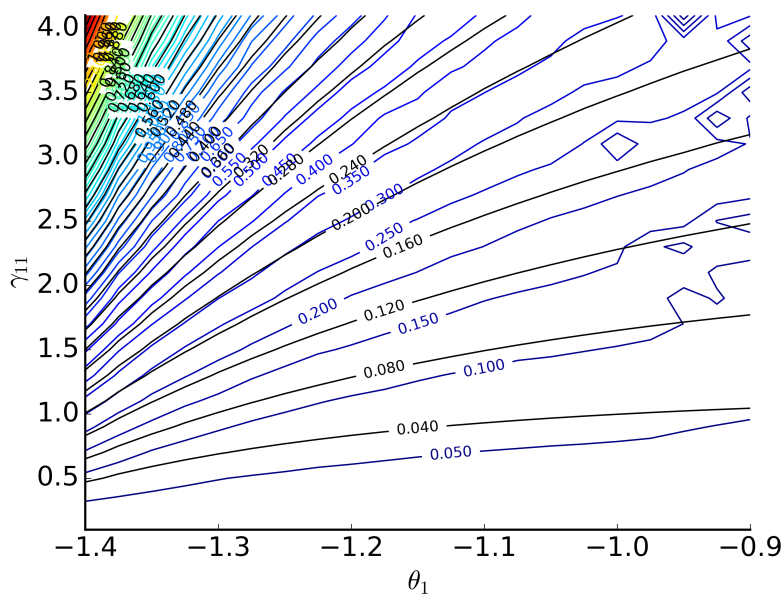

**Fig. S3.** Comparison of the speed of propagation obtained by the shooting method (black lines) and the simulation of the full model (colored lines).

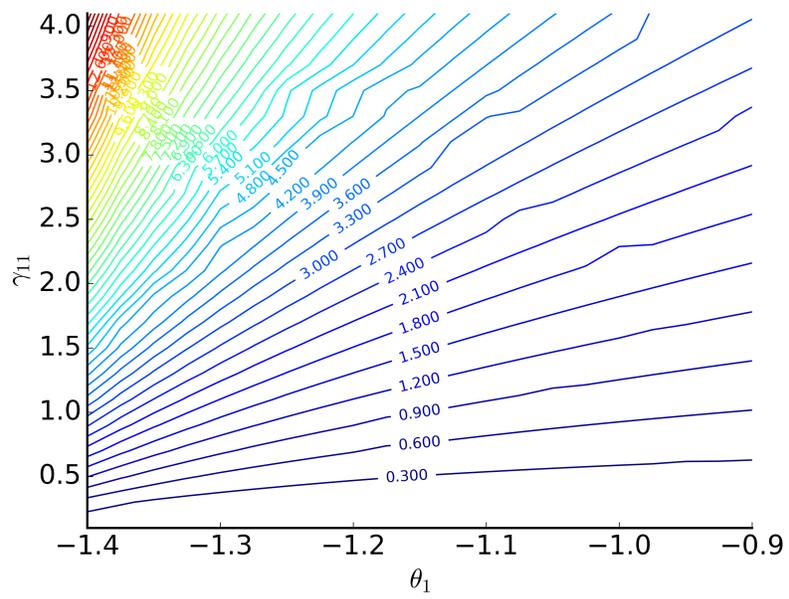

**Fig. S4.** Shooting method for the fast population in the reduced model.

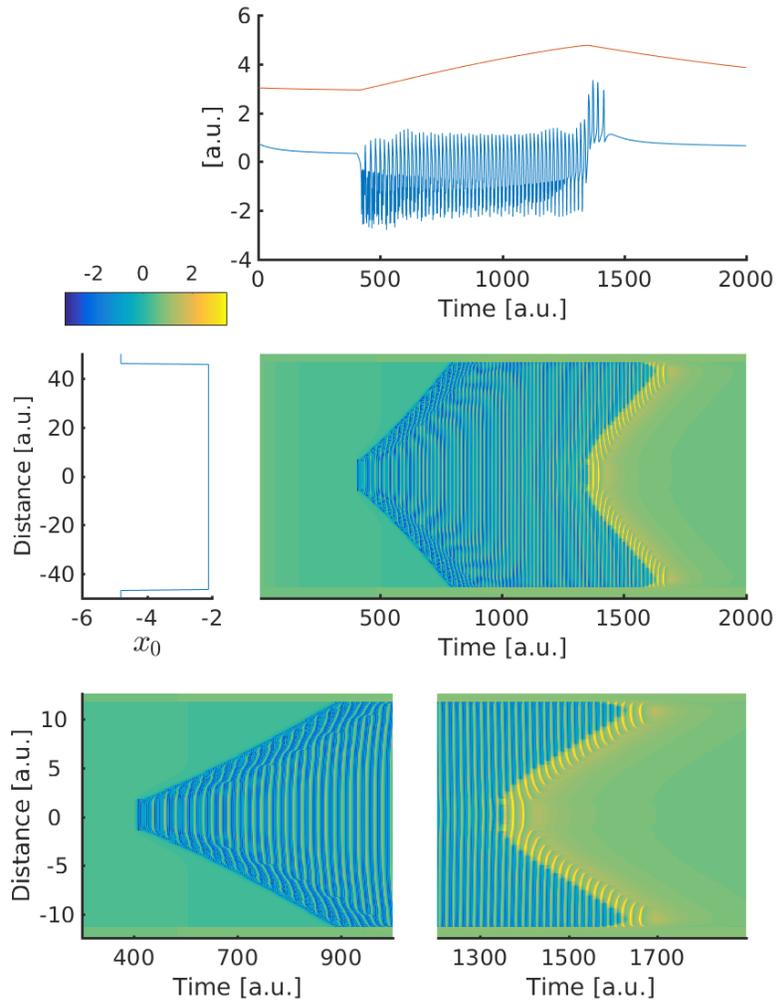

**Fig. S5.** Seizures do not end synchronously across the field if the delays of recruitment are too big compared to the length of the seizure. Large delays were here obtained by increasing the spatial domain. A single Epileptor field model is simulated (middle row). Spatial values for excitability parameter and simulated spatiotemporal activity are shown on the left and right middle row respectively. The seizure starts in the central location of the Epileptor field model, and slowly propagates throughout the field. Time series at the zero location distance (where the seizure first starts) are shown in the top row for the field activity ($-u_1 + q_1$, in blue) and the slow permittivity variable ($s$, in red). The lower row shows a zoomed-in view of the spatiotemporal activity at seizure onset (left) and offset (right). The seizure do not ends synchronously in the field, with a few spike-and-wave events propagating after seizure offset.

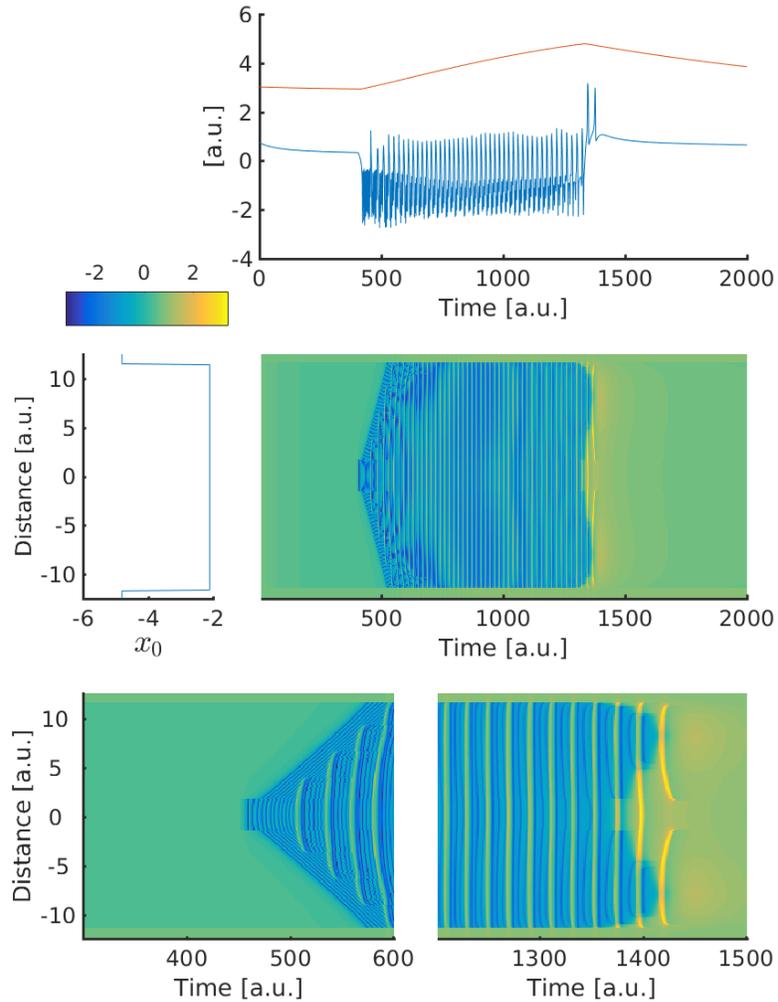

**Fig. S6.** The number of SWDs after seizure termination changes as a function of the strength of the coupling function $\gamma_{12}$ ($\gamma_{12} = 0.5$). A single Epileptor field is simulated (middle row). Spatial values for excitability parameter and simulated spatiotemporal activity are shown on the left and right middle row respectively. The seizure starts in the central location of the Epileptor field model, and slowly propagates throughout the field. Time series at the zero location (where the seizure first starts) are shown in the top row for the field activity ($-u_1 + q_1$, in blue) and the slow permittivity variable ($s$, in red). The lower row shows a zoomed-in view of the spatiotemporal activity at seizure onset (left) and offset (right). Only two SWDs appear after seizure offset.

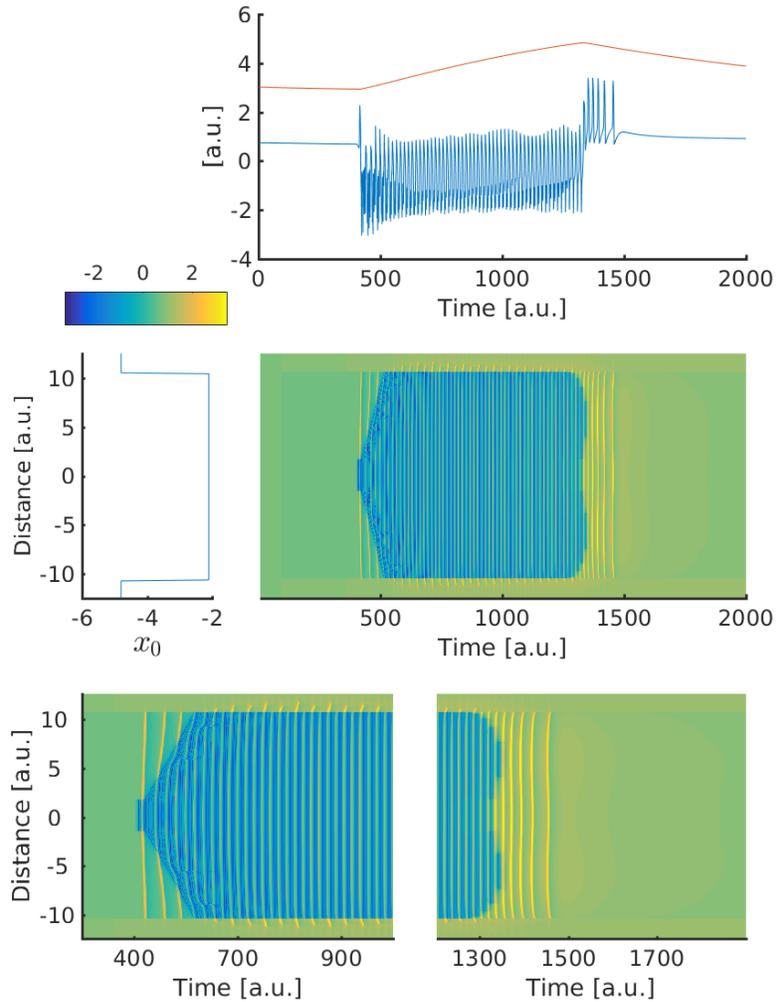

**Fig. S7.** Spike-and-wave discharges are shifted towards the beginning of the seizure when changing the parameter $\gamma$ of the temporal convolution ($\gamma = 0.01$). A single Epileptor fields is simulated (middle row). Spatial values for excitability parameter and simulated spatiotemporal activity are shown on the left and right middle row respectively. The seizure starts in the central location of the Epileptor field model, and slowly propagates throughout the field. Time series at the zero location (where the seizure first starts) are shown in the top row for the field activity ($-u_1 + q_1$, in blue) and the slow permittivity variable ($s$, in red). The lower row shows a zoomed-in view of the spatiotemporal activity at seizure onset (left) and offset (right). Preictal spikes appear before seizure onset, propagating toward easily excitable regions.

**Proix et al.**